\documentclass[prb]{revtex4}
   
\usepackage{amsmath}
\usepackage{graphicx}
\usepackage{color}

\begin{document}

\title{Prediction of Debye-Scherrer diffraction patterns in arbitrarily strained samples}
\author{Andrew Higginbotham}
\affiliation{Department of Physics, Clarendon Laboratory, University of Oxford, Parks Road, Oxford OX1 3PU, UK}
\author{David McGonegle}
\affiliation{Department of Physics, Clarendon Laboratory, University of Oxford, Parks Road, Oxford OX1 3PU, UK}

%\date{\today}

\begin{abstract}
The prediction of Debye-Scherrer diffraction patterns from strained samples is typically conducted in the small strain limit.  Although valid for small deviations from the hydrostat (such as the conditions of finite strength typically observed in diamond anvil cells) this assertion is likely to fail for the large strain anisotropies (often of order 10\% in normal strain) such as those found in uniaixally loaded dynamic compression experiments.  In this paper we derive a general form for the $\left( \theta_B, \phi\right)$ dependence of the diffraction for an arbitrarily deformed sample in arbitrary geometry.  We show that this formula is consistent with ray traced diffraction for highly strained computationally generated polycrystals, and that the formula shows deviations from the small strain solutions previously reported.
\end{abstract}

\maketitle

\section{Introduction}

Powder diffraction has been used as a powerful diagnostic of crystal structure for many decades, and remains one of the most sensitive techniques we have for probing the structure of materials.  In particular, a plethora of complex high pressure phases have been discovered in a diverse range of material \cite{Gregoryanz2008,Loa2012,McMahon2000}.  A nearly universal trend has been seen, with materials becoming complex as pressure rises, in direct contradiction with the traditional view that solids at high pressure will metallize and collapse down to a simple close packed structure.  This onset of complexity of pressures above 100\,GPa is particularly intriguing in light of the increased interest in potential planetary constituent materials given the recent confirmed discovery of over 800 exoplanets, where current models suggest a broard range of planetary classes \cite{Swift2012}.\\
With pressures of interest reaching several TPa, traditional static high pressure techniques, where samples at near hydrostatic conditions are readily attained, are inadequate.  For example, recent work by Rygg and coworkers has demonstrated powder diffraction from solid samples at pressures of up to 800GPa \cite{Rygg2012}, with pressure of over 1\,TPa now readily achievable on large scale laser facilities \cite{G.W.Collins}.    However, in order to reach such conditions, transient laser compression is employed.   In this technique uniaxial shock, or for higher pressures, ramp waves are launched into the sample leading to an initial state of uniaxial elastic strain.  In addition, enhanced strength of materials at high pressure could potentially lead to large departures from hydrostatic response, even after plasticity or structural phase change have relieved the initial elastic strain anisotropy.   This departure from traditional small deviatoric strain limit leads to a need for more careful analysis of the response of Debye-Scherrer rings from material under compression.  In this paper we detail a general formula for the deflection of a Debye-Scherrer ring in $\theta$ which allows for $\phi$ dependence (a consequence of certain geometries where the incident X-ray direction is not the same as normal strain direction) and arbitrary deformation within the Voigt limit.  This formula will be compared with the more traditional approach for small strains developed by Singh \cite{singh1993lss}, and with simulated Debye-Scherrer patterns from molecular dynamics simulations. \\

\section{Derivation}

\label{sec:derivation}
\begin{figure}
\begin{center}
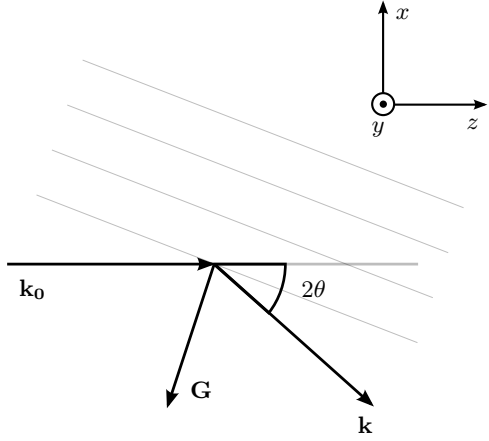
\caption{Schematic of the working coordinates for the derivation.}
\label{default}
\end{center}
\end{figure}
For simplicity, and without loss of generality, we work in a coordinate system where the incident X-ray direction, $\mathbf{k_0}$, is along $z$.  Note that in general this choice clashes with the convention that $z$ is the shock direction in the sample. This leads us to define two coordinate systems, one attached to the sample, where the shock is along $z$, and one, referred to as working coordinates, where the X-ray incidence direction defines $z$.  We also define a rotation matrix, $\mathbf{R}$, which transforms us from working coordinates to sample coordinates.  Note that for X-rays incident along the shock direction $\mathbf{R}=\mathbf{I}$, the identity matrix.\\
To start, we note that any general diffracting plane will have outgoing $\mathbf{k}$ vector defined, in the working coordinate system, by -
\begin{align}
\mathbf{k}= \frac{2\pi}{\lambda}\left( \begin{array}{c} \sin{2\theta_B}\cos{\phi} \\ \sin{2\theta_B}\sin{\phi} \\ \cos{2\theta_B}\end{array}\right)
\end{align}
where $\theta_B$ is the Bragg angle \emph{in the compressed system} and $\phi$ the azimuthal angle around the Debye-Scherrer ring. By noting the Laue condition ($\mathbf{G}=\mathbf{k}-\mathbf{k_0}$) we define $\mathbf{G}$, the diffracting reciprocal lattice vector, as -
\begin{align}
\mathbf{G}= \frac{2\pi}{\lambda}\left( \begin{array}{c} \sin{2\theta_B}\cos{\phi} \\ \sin{2\theta_B}\sin{\phi} \\ \cos{2\theta_B} -1\end{array}\right)
\end{align}
Note that this reciprocal lattice vector denotes a set of planes, within a certain grain in the sample, which now meet the Bragg condition after being deformed.  They are not, in general, the same planes which diffracted in the undeformed sample (i.e. as we compress a sample, we actually change the subset of grains which we are probing).\\
We proceed by finding the equivalent vector, $\mathbf{G_0}$ in the undeformed system.  We limit discussion here to the Voigt (iso-strain) condition, under which the deformation of all grains can be defined by a single deformation gradient, $\mathbf{F}$.  Since $\mathbf{G}$ is a reciprocal lattice vector we must consider not the deformation in real space, but the associated `deformation' of reciprocal space.  For a general deformation gradient in real space the reciprocal space analogue is given by $\cal F\rm = \left(\mathbf{F}^T\right)^{-1}$ (see appendix \ref{app:deformation}). Thus, to return the deformed reciprocal space back to the undeformed we use $\cal F \rm^{-1}  = \mathbf{F}^{T}$.\\  
This gives $\mathbf{G_0}$, the original, unstrained reciprocal lattice vector as -

\begin{align}
\mathbf{G_0}^\prime & = \mathbf{F^TRG} = \mathbf{\boldsymbol\alpha G}  \label{eg:transform}\\
&= \frac{2\pi}{\lambda} \left( \begin{array}{c c c}
\alpha_{11} & \alpha_{12} & \alpha_{13} \\
\alpha_{21} & \alpha_{22} & \alpha_{23} \\
\alpha_{31} & \alpha_{32} & \alpha_{33} 
\end{array} \right)
\left( \begin{array}{c} \sin{2\theta_B}\cos{\phi} \\ \sin{2\theta_B}\sin{\phi} \\ \cos{2\theta_B} -1 \end{array}\right)\\
&=\frac{2\pi}{\lambda} \left( \begin{array}{c} 
\alpha_{11}\sin{2\theta_B}\cos{\phi}  + \alpha_{12}\sin{2\theta_B}\sin{\phi}  + \alpha_{13}\left( \cos{2\theta_B} -1\right)\\ 
\alpha_{21}\sin{2\theta_B}\cos{\phi}  + \alpha_{22}\sin{2\theta_B}\sin{\phi}  + \alpha_{23}\left( \cos{2\theta_B} -1\right)\\ 
\alpha_{31}\sin{2\theta_B}\cos{\phi}  + \alpha_{32}\sin{2\theta_B}\sin{\phi}  + \alpha_{33}\left( \cos{2\theta_B} -1\right)\\
\end{array}\right)
\end{align}

where $\mathbf{\alpha}=\mathbf{F^TR}$.  The prime denotes that the reciprocal lattice vector is expressed in terms of sample coordinates.  One could of course apply the transpose rotation tensor, $\mathbf{R^T}$, from the left hand side in equation \ref{eg:transform}, rotating the result back in to the working coordinates.  However, this would complicate the form of $\boldsymbol\alpha$, and as we will see, the direction of $\mathbf{G_0}$ is of no consequence.\\ 
As noted above, the fact that $\mathbf{G}$ meets the Laue condition is no guarantee that $\mathbf{G_0}$ did, so the only information of use in $\mathbf{G_0}$ is its length, $|\mathbf{G_0}^\prime|=\frac{2\pi}{d_0}$.  By exploiting this knowledge one can assign a value to $|\mathbf{G_0}|^2$ and thus recover and expression for linking $\theta_B$ and $\phi$ such that -

\begin{align}
\frac{\lambda^2}{d_0^2} &= \left(\alpha_{11}^2 + \alpha_{21}^2 + \alpha_{31}^2\right)\sin^2{2\theta_B}\cos^2{\phi}  \notag\\
&+   2\left(\alpha_{11} \alpha_{12} + \alpha_{21} \alpha_{22} + \alpha_{31} \alpha_{32}\right)\sin^2{2\theta_B}\cos{\phi}\sin{\phi} \notag\\
&+   2\left(\alpha_{11} \alpha_{13} + \alpha_{21} \alpha_{23} + \alpha_{31} \alpha_{33}\right) \sin{2\theta_B}\cos{\phi}\left( \cos{2\theta_B} -1\right) \notag\\
&+   \left(\alpha_{12}^2 + \alpha_{22}^2 + \alpha_{32}^2\right) \sin^2{2\theta_B}\sin^2{\phi}  \notag\\
&+   2\left(\alpha_{12} \alpha_{13} + \alpha_{22} \alpha_{23} + \alpha_{32} \alpha_{33}\right)  \sin{2\theta_B}\sin{\phi}\left( \cos{2\theta_B} -1\right) \notag\\
&+   \left(\alpha_{13}^2 + \alpha_{23}^2 + \alpha_{33}^2\right) \left( \cos{2\theta_B} -1\right)^2\\
\label{eq:final}
&= \sin^2{2\theta_B}\left(A_1\cos^2{\phi} + 2A_2 \cos{\phi}\sin{\phi} + A_4\sin^2{\phi}\right) \notag\\
&+  \sin{2\theta_B}\left( \cos{2\theta_B} -1\right) \left(  2A_3\cos{\phi} + 2A_5\sin{\phi} \right) \notag\\
&+ \left( \cos{2\theta_B} -1\right)^2A_6
\end{align}

where the $A$ coefficients correspond to the combinations of rotated deformation gradient components.  This equation gives the relation between $\theta_B$ and $\phi$ as one proceeds around the Debye-Scherrer ring.  

\section{Solutions}

\subsection{$\mathbf{R}=\mathbf{I}$, normal strain only}

In this geometry we have the sample coordinate system identical to our working geometry such that $\mathbf{R}=\mathbf{I}$.  We also assume that the off-diagonal elements of the strain tensor are zero (as we typically do not consider pure shear in shock physics applications).  In this case equation \ref{eq:final} can be simplified by noting that the only non-zero coefficients are $A_1=\alpha^2_{11}$, $A_4=\alpha^2_{22}$ and $A_6=\alpha^2_{33}$ so that --

\begin{align}
\frac{\lambda^2}{d_0^2} &=   \sin^2{2\theta_B}\left( \alpha^2_{11}\cos^2{\phi} + \alpha^2_{22}\sin^2{\phi}  \right) +\notag \\
 & \left( \cos{2\theta_B} -1\right)^2\alpha^2_{33}
\label{eq:Hull}
\end{align}
Note that in this case the rotated deformation gradient is simply --
\begin{align}
\mathbf{\alpha} = \left( \begin{array}{c c c}
1+\varepsilon_{xx} & 0 & 0 \\
0 &1+\varepsilon_{yy} & 0 \\
0 & 0 &1+\varepsilon_{zz} 
\end{array}\right)
\end{align}
Rearranging equation \ref{eq:Hull} we arrive at --
\begin{align}
\sin^4\theta_B \left(  \alpha_{33}^2 -  \alpha^2_{11}\cos^2{\phi} - \alpha^2_{22}\sin^2{\phi} \right) + \notag \\
\sin^2\theta_B\left( \alpha^2_{11}\cos^2{\phi} + \alpha^2_{22}\sin^2{\phi}  \right) - \frac{\lambda^2}{4d_0^2} = 0
\end{align}
For $\alpha_{11}=\alpha_{22}=\alpha_{33}$ we recover hydrostatic compression and the expression is seen to reduce to  Bragg's law as expected.

\subsection{Tilted target geometry}
\label{sec:tilted}

Although the $\mathbf{R}=\mathbf{I}$ geometry discussed above is common in static experiments, shock physics environments rarely allow such symmetry.  X-ray sources can be difficult to place along the loading axis, and as we typically only compress the central portion of the sample we can not probe at purely transverse incidence. In addition, the non normal X-ray incidence can be of some advantage.  As shown by Singh, and in the shock case by Hawreliak and coworkers, by tilting the sample relative to the incoming X-rays we encode information on differing strain components around the Debye-Scherrer ring \cite{singh1993lss,Hawreliak2007,Hawreliak2011}.\\   
In tilted target geometry we apply a rotation of $\chi$ about $y$ to give --
\begin{align}
\mathbf{\alpha} &=
 \left( \begin{array}{c c c}
1+\varepsilon_{xx} & 0 & 0 \\
0 &1+\varepsilon_{yy} & 0 \\
0 & 0 &1+\varepsilon_{zz} 
\end{array}\right)
 \left( \begin{array}{c c c}
\cos{\chi} & 0 & -\sin{\chi} \\
0 &1 & 0 \\
\sin{\chi} & 0 &\cos{\chi}
\end{array}\right)\\
&= \left( \begin{array}{c c c}
\left(1+\varepsilon_{xx}\right)\cos{\chi} & 0 & -\left(1+\varepsilon_{xx}\right)\sin{\chi} \\
0 &1+\varepsilon_{yy} & 0 \\
\left(1+\varepsilon_{zz}\right)\sin{\chi} & 0 & \left(1+\varepsilon_{zz}\right)\cos{\chi} 
\end{array}\right)
\end{align}\\
where we have once again assumed only normal strains for simplicity.  This gives --

\begin{align}
\frac{\lambda^2}{d_0^2} &= \left( \left(1+\varepsilon_{xx}\right)^2\cos{\chi}^2\cos^2{\phi} \notag  + \left(1+\varepsilon_{yy}\right)^2 \sin^2{\phi} + \left(1+\varepsilon_{zz}^2\right)\sin^2{\chi}\cos^2{\phi}\right) \sin^2{2\theta_B}\notag\\
& + 2\cos{\chi}\sin{\chi}\left( (1+\varepsilon_{zz})^2 - (1+\varepsilon_{xx})^2 \right) \cos{\phi}\sin{2\theta_B}\left( \cos{2\theta_B}-1\right)\notag \\
& + \left( \left( 1+\varepsilon_{xx}\right)^2\sin^2\chi +  \left(1+\varepsilon_{zz}\right)^2\cos^2\chi   \right)\left( \cos{2\theta_B}-1\right)^2  \label{eq:tiltedFull}
\end{align}

Once again, we note that this reduces to the $\mathbf{R}=\mathbf{I}$ geometry equation for $\chi=0$.  In order to solve this equation for $\theta$ we define deformation, $\phi$ and $\chi$ coefficients such that -
\begingroup
\allowdisplaybreaks

\begin{align}
D_1  \sin^2{2\theta_B} + D_2\sin{2\theta_B}\left( \cos{2\theta_B}-1\right) + D_3 \left( \cos{2\theta_B}-1\right)^2 -\frac{\lambda^2}{d_0^2}&=0 \label{eq:tiltedWithD}\\
D_1 \sin^2{2\theta_B} - 2D_2\sin{2\theta_B}\sin^2{\theta_B} + 4D_3 \sin^4{\theta_B}    -\frac{\lambda^2}{d_0^2}&=0\\
-D_1 \left( e^{i2\theta_B} - e^{-i2\theta_B}\right)^2 - iD_2  \left( e^{i2\theta_B} - e^{-i2\theta_B}\right)\left( e^{i\theta_B} - e^{-i\theta_B}\right)^2\notag\\ 
+D_3\left( e^{i\theta_B} - e^{-i\theta_B}\right)^4 - \frac{4\lambda^2}{d_0^2} &= 0\\
-D_1\left( e^{i4\theta_B} + e^{-i4\theta_B} -2  \right) - iD_2\left( e^{i4\theta_B} - e^{-i4\theta_B} -2e^{i2\theta_B} +2e^{-i2\theta_B}\right)\notag \\
+D_3\left(  e^{i4\theta_B} + e^{-i4\theta_B} - 4e^{i2\theta_B} - 4e^{-i2\theta_B} +6 \right)  - \frac{4\lambda^2}{d_0^2} &= 0\\
-D_1\left( e^{i8\theta_B} + 1 -2e^{i4\theta_B}  \right) - iD_2\left( e^{i8\theta_B} - 1 -2e^{i6\theta_B} +2e^{i2\theta_B}\right)\notag \\
+D_3\left(  e^{i8\theta_B} + 1 - 4e^{i6\theta_B} - 4e^{i2\theta_B} +6e^{i4\theta_B} \right)  - \frac{4\lambda^2}{d_0^2}e^{i4\theta_B} &= 0 \\
\left(  D_3 -D_1 -iD_2 \right)e^{i8\theta_B}  - \left(4D_3 - 2iD_2  \right)e^{i6\theta_B} + \left(2D_1 + 6D_3 -     \frac{4\lambda^2}{d_0^2} \right) e^{i4\theta_B} \notag \\
-\left(2iD_2 + 4D_3 \right)e^{i2\theta_B} + \left(  D_3 + iD_2 -D_1 \right)=0 \label{eq:tiltedFinal}
\end{align}
\endgroup

\noindent where the D coefficients are defined by comparison between equations \ref{eq:tiltedFull} and \ref{eq:tiltedWithD}.  This quartic in $e^{i2\theta_B}$ can be solved to find $\theta_B$.

\subsection{The general solution}
It should be noted that the derivation of section \ref{sec:tilted} is equally applicable in the general case (though the normal strain, tilted target solution is expected to be sufficient for most applications).  Here, we refer to equation \ref{eq:final} to see that we can define the D coefficients of equation \ref{eq:tiltedWithD} in a more general sense as --
\begin{align}
D_1 &= A_1\cos^2{\phi} + 2A_2 \cos{\phi}\sin{\phi} + A_4\sin^2{\phi} \\
D_2 &=  2A_3\cos{\phi} + 2A_5\sin{\phi}\\
D_3 &=  A_6
\end{align}
The solution in equation \ref{eq:tiltedFinal} is then equally valid with these new D coefficients.  A summary of the full form of the $A_i$ and $D_i$ coefficients is given in table \ref{tab:coeffs}
\begin{table}
\caption{Summary of the $A_i$ and $D_i$ coefficients in terms of the $\alpha_{ij}$ components.}
\begin{center}
\begin{tabular}{cc|cl}
$A_1$ & \quad \quad&  \quad \quad & $\alpha_{11}^2 + \alpha_{21}^2 + \alpha_{31}^2$ \\
$A_2$ & \quad \quad&  \quad \quad & $\alpha_{11} \alpha_{12} + \alpha_{21} \alpha_{22} + \alpha_{31} \alpha_{32}$ \\
$A_3$ & \quad \quad&  \quad \quad &  $\alpha_{11} \alpha_{13} + \alpha_{21} \alpha_{23} + \alpha_{31} \alpha_{33}$ \\
$A_4$ & \quad \quad&  \quad \quad & $\alpha_{12}^2 + \alpha_{22}^2 + \alpha_{32}^2$ \\
$A_5$ & \quad \quad&  \quad \quad & $\alpha_{12} \alpha_{13} + \alpha_{22} \alpha_{23} + \alpha_{32} \alpha_{33}$\\
$A_6$ & \quad \quad&  \quad \quad & $\alpha_{13}^2 + \alpha_{23}^2 + \alpha_{33}^2$\\
& &\\
\hline
& &\\
$D_1$ & \quad \quad&  \quad \quad & $A_1\cos^2{\phi} + 2A_2 \cos{\phi}\sin{\phi} + A_4\sin^2{\phi}$  \\
$D_2$ & \quad \quad&  \quad \quad & $2A_3\cos{\phi} + 2A_5\sin{\phi}$ \\
$D_3$ & \quad \quad&  \quad \quad & $A_6$
\end{tabular}
\end{center}
\label{tab:coeffs}
\end{table}%
\section{Verification}
\label{sec:verification}

\begin{figure}
\begin{center}
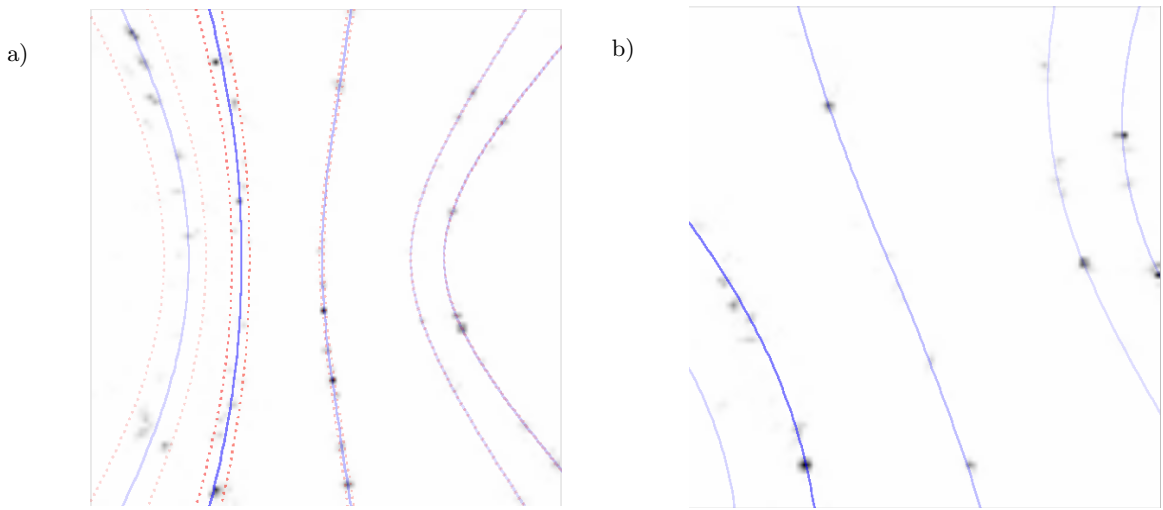
\caption{Raytraced diffraction signals from deformed polycrystals.  Theory overlays are shown as (solid) blue lines.  Panel a shows a detector 400x400mm, 10mm away from the sample with the centre of the detector corresponding to $2\theta_B=90^\circ$, $\phi=\pi$, in a geometry with $\mathbf{R}=\mathbf{I}$ and the deformation defined by equation \ref{eq:DeformationA}.  Panel b shows the same detector at 283mm away, again at $2\theta_B=90^\circ$, $\phi=\pi$, with $\chi=\pi/4$ and the deformation defined by equation \ref{eq:DeformationB}.}
\label{fig:raytrace}
\end{center}
\end{figure}

In order to verify the previous formulae we compare with synthetic diffraction data obtained by raytracing from polycrystals.  A $300\times300\times300$\,\AA\  cube of material with $300$ randomly oriented grains was computationally generated and deformed according to a specified deformation gradient \cite{Traiviratana2008}.  The resultant set of atomic coordinates was used as the input to a raytracing calculation in a variety of geometries \cite{Kimminau2008}.  In this calculation, `photons' are traced from detector pixel back to source, via a sample position.  in doing so, for a given photon wavelength we define, via the Laue condition, a diffracting G-vector.  The diffraction signal is simply related to the intensity of the Fourier transform of atomic positions -
\begin{align}
I\left(G\right) = \left| \sum_j e^{i\mathbf{G.r_j}} \right| ^2
\end{align}
where the sum is over all atoms in the diffracting sample.  Having defined an intensity per pixel we have an artificial diffraction image, which should show deformation of the Debye-Scherrer rings consistent with the theory above provided a consistent use of the deformation gradient.\\
To demonstrate the agreement between theory and raytrace we show two examples.  One, for $\mathbf{R}=\mathbf{I}$ and normal strain only corresponds to a common synchrotron geometry (where, for example, small departures from hydrostatic stress are seen in diamond anvil cells).  In this case we choose a simple deformation gradient of the form -
\begin{align}
\mathbf{F_a}  = \mathbf{\alpha} = \left( \begin{array}{c c c}
0.8 & 0.0 & 0.0 \\
0.0 & 0.9 & 0.0 \\
0.0 & 0.0 & 0.7
\end{array}\right)
\label{eq:DeformationA}
\end{align}  
This deformation gradient is applied to the atomic positions of the initially undeformed polycrystal.  Raytracing of 9\,keV photons to a detector plane with normal perpendicular to the load axis results in figure \ref{fig:raytrace}a.  The simulated diffraction from grains which meet the Bragg condition in the deformed sample match well with the theory above (blue lines).  Also shown are theory lines for deformation gradients where we use $\alpha_{33}=0.7\pm0.01$, corresponding to around a 1.5\% change in volume of the deformed sample.  It should be noted that diffraction from all grains lies within these bounds, and that the bounds reflect the symmetry of the grain distribution, indicative of the blue theory lines lying at the centre of the Debye-Scherrer ring.\\
A more complex example is shown in figure \ref{fig:raytrace}b.  In this case we use a rotation of $\chi=\pi/4$ and a deformation gradient of --
\begin{align}
\mathbf{F_b}  = \mathbf{\alpha} = \left( \begin{array}{c c c}
0.8 & 0.12 & -0.07 \\
0.13 & 0.9 & 0.15 \\
0.04 & -0.04 & 0.7
\end{array}\right)
\label{eq:DeformationB}
\end{align}  
This geometry has a detector with normal perpendicular to $\mathbf{k_0}$, and a photon energy of 8\,keV.  Again, good agreement between ray trace and theory is seen, verifying the form of equation \ref{eq:tiltedFull}.\\

\section{Strength measurement}

One key consideration in shock compression is the degree of departure from hydrostatic stress.  This property, known as strength, may exhibit itself in the finite degree of initial uniaxial elastic compression (i.e. yield strength, or Hugoniot Elastic Limit) or in residual elastic strain anisotropy after plastic deformation or phase change.  Similar scenarios have been studied in terms of non-hydrostatic conditions between the planar culets of conventional diamond anvil cells.    For example, Singh discusses the effects of finite shear stress on the diffraction from high pressure samples \cite{singh1993lss}.  As one of the few extant theories concerning diffraction in highly non-hydrostatic conditions, we compare our results to the widely used formulae of Singh.\\
We start with the outgoing $\mathbf{G}$ (noting it has a length of $\frac{2\pi}{d}$) --
\begin{align}
\mathbf{G} = \frac{2\pi}{d}\left( \begin{array}{c} \sin\psi \\ 0 \\\cos\psi \end{array}\right)
\end{align}
where $\psi$ is the angle between $\mathbf{G}$ and the compression direction.  As before, we work out the equivalent vector, $\mathbf{G_0}$, in the undeformed reciprocal lattice.  Thus we apply a deformation gradient (and since rotations do not affect the result, an $\alpha$ tensor of) --
\begin{align}
\mathbf{F^T} = \mathbf{\alpha} = 
\left(
\begin{array}{ccc}
1+\varepsilon_p-\gamma/3 & 0 & 0 \\
0 & 1+\varepsilon_p-\gamma/3 & 0 \\
0 & 0 & 1+\varepsilon_p+2\gamma/3 \\
\end{array}
\right)
\end{align}
where, following Singh,  $\epsilon_p=\frac{1}{3}\left(2\varepsilon_{xx}+\varepsilon_{zz}\right)$ and $\gamma=\varepsilon_{zz}-\varepsilon_{xx}$ is the shear strain.  Applying this to $\mathbf{G}$, and taking the modulus squared --
\begin{align}
\left( \frac{2\pi}{d_0} \right)^2 &= \left( \frac{2\pi}{d} \right)^2 \left( \alpha_{11}^2\sin^2\psi + \alpha_{33}^2\cos^2\psi \right) \\
\Rightarrow \left( \frac{d}{d_0} \right)^2 &=  \alpha_{11}^2\left( 1 -\cos^2\psi\right) + \alpha_{33}^2\cos^2\psi \\
&=  \alpha_{11}^2 + \left(\alpha_{33}^2-\alpha_{11}^2\right)\cos^2\psi 
\label {eq:largeStrain}
\end{align}
Since Singh works exclusively in the small $\gamma$ limit we note that --
\begin{align}
\alpha_{11}^2 &\approx \left(1+\varepsilon_p\right)^2 - 2\gamma\left(1+\varepsilon_p\right)/3 \\
\alpha_{33}^2 &\approx \left(1+\varepsilon_p\right)^2 + 4\gamma\left(1+\varepsilon_p\right)/3 
\end{align}
where $\varepsilon_d^V$ is the change in d-spacing due to the deviatoric strain component in the Voigt limit.  This allows us to express equation \ref{eq:largeStrain} as --
\begin{align}
\left(\frac{d}{d_0}\right)^2  &\approx \left(1+\varepsilon_p\right)^2 - \frac{2\left(1+\varepsilon_p\right)\gamma}{3}\left(1-3\cos^2\psi\right) \\
&\approx \left ( \left(1+\varepsilon_p\right) - \frac{\gamma}{3}\left( 1-3\cos^2\psi\right) \right)^2\\
\Rightarrow \frac{d}{d_0} -1 &=  \varepsilon_p + \varepsilon_d^V \approx  \varepsilon_p - \frac{\gamma}{3}\left( 1-3\cos^2\psi\right) 
\end{align}  
This small shear strain formula is in agreement with Singh's result.  However, we note that equation \ref{eq:largeStrain} forms a more general relation between the measurable quantity, $d/d_0$ and the angle of $\mathbf{G}$ relative to loading, which does not assume small strains.\\  

\begin{figure}
\begin{center}
\caption{Comparison of the theory of section \ref{sec:derivation} with that of Singh. Black lines represent this paper's theory, with the red line in the left panel showing the small strain theory predictions for the same deformation.}
\includegraphics[width=12cm]{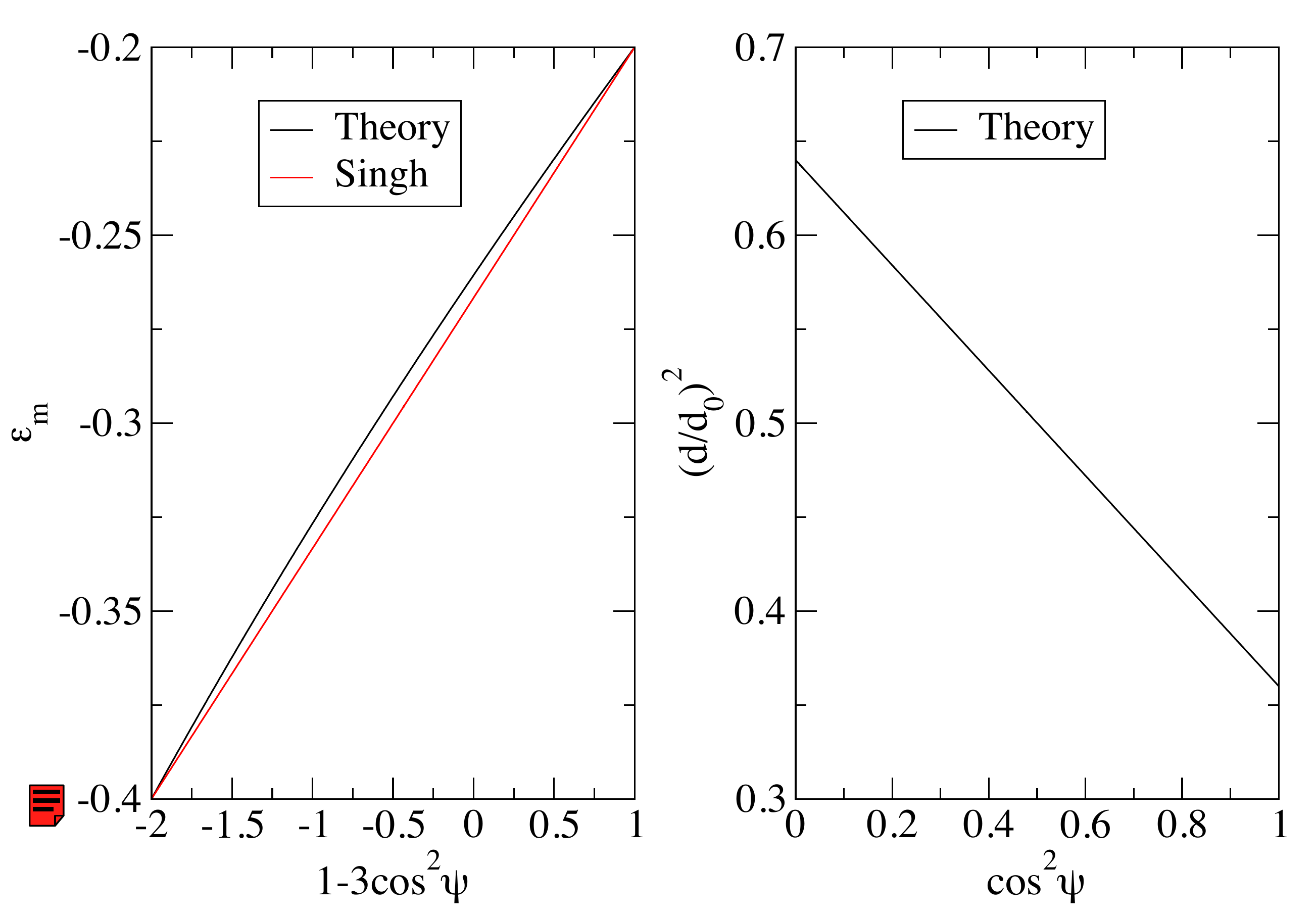}
\label{fig:shear}
\end{center}
\end{figure}
 
To demonstrate the deviation expected from Singh's formula for large shear we show the plots based on Singh's methodology as well one based on equation \ref{eq:largeStrain}.  In Singh's formalism it is noted a of a plot of $1-3\cos^2\psi$ against $\varepsilon_m = \left(d-d_0\right)/d_0$, the measured strain, is linear with gradient $\gamma/3$.  Although the theory presented above agrees well with this at low $\gamma$, deviations are typically seen for shear strains of 10\% and above, a level of shear which can be seen in the elastic compression phase of a shock \cite{Whitley2011}.  It should be noted that even for the case of $\varepsilon_p=-0.4$, $\gamma=-0.2$ shown in figure \ref{fig:shear}, departures of the current theory from Singh's linear equation are minimal.  However, if only a small range of $\cos^2\psi$ is accessible experimentally, a linear fit may lead to significant errors of several percent in strain, both volumetric and shear.\\
The proposed $\cos^2\psi$ against $\left(d/d_0\right)^2$ plot is seen to be linear over all $\cos^2\psi$, making it a better candidate for shear strain (and isotropic strain) determination where limited $\psi$ range is covered.\\
In the case of non-zero off-diagonal deformation gradient components both of these analyses fail, as measured strain is no longer single-valued in $\psi$.  Here, a full fit to the diffraction pattern using equation \ref{eq:tiltedFull} is required to determine the applied deformation.
 
\section{Conclusions} 

We derive an expression for the form of Debye-Scherrer diffraction from samples with arbitrary deformation, and with arbitrary orientation of loading directions with respect to incident X-ray direction.  This expression is shown to agree with raytracing from computationally generated polycrystals, and in the small strain limit with existing theory.  However, it is shown that for large shear strain, a more accurate form of analysis can alleviate errors due to incomplete sampling of $\psi$, and that this form is applicable for any strain state within the Voigt limit.\\
As noted above, deviations from existing small strain theory start at around $\gamma=0.05$, and are pronounced above $\gamma=0.1$.  Errors of the order of percent in strain can be introduced by assuming a linear, small strain response in the $1-3\cos^2\psi$ formalism. However, a small modification to the analysis removes this source of error without adding complexity.\\

\section*{Acknowledgements}

The authors would like to thank M. J. Suggit and J. S. Wark  for fruitful discussions and useful comments. The authors are grateful for support from AWE and LLNL under sub-contract B595954..

\appendix

\section{Reciprocal space deformation}

\label{app:deformation} 
Let us assume a deformation gradient, $\mathbf{F}$, is applied in real space. This deformation gradient is defined such that the position, $\mathbf{u}$, of an element in the undeformed system,  is related to its position in the deformed system, $\mathbf{U}$, by $\mathbf{U}=\mathbf{Fu}$.  In general this deformation gradient will consist of 9 independent components -
\begin{align}
\mathbf{F} = \left( \begin{array}{ccc} 
a & b & c\\
d & e & f\\
g & h & i
\end{array} \right)
\end{align}  
One can apply this to three unit vectors initially aligned with the cartesian axes in the undeformed system to fully characterise the deformation -
\begin{align}
\mathbf{a}  = \left( \begin{array}{c} a\\ d\\ g \end{array} \right), \quad \mathbf{b}  = \left( \begin{array}{c} b\\ e\\ h \end{array} \right), \quad \mathbf{c}  = \left( \begin{array}{c} c\\ f\\ i\end{array} \right)
\end{align}
Assuming that these vectors span a lattice, one can define the equivalent reciprocal lattice vectors as -
\begin{align}
\mathbf{a^\ast} = \frac{2\pi}{V_D} \left( \begin{array}{c} ei-hf\\ hc-bi\\ bf-ce \end{array} \right)\\
\mathbf{b^\ast} = \frac{2\pi}{V_D} \left( \begin{array}{c} fg-id\\ ia-cg\\ cd-af \end{array} \right)\\
\mathbf{c^\ast} = \frac{2\pi}{V_D} \left( \begin{array}{c} dh-eg\\ bg-ah\\ ae-db \end{array} \right)
\end{align}
where $V_D=a\left(ei-hf\right) + d\left(hc-bi\right) + g\left(bf-ce\right)$ is the volume of an initially cubic unit cell in the deformed system.  One can use these reciprocal lattice vectors to define a deformation gradient, $\cal{F}$, for reciprocal space -
\begin {align}
\cal{F} = \rm \frac{1}{V_D} \left( \begin{array}{ccc} 
 ei-hf& fg-id\ &   dh-eg\\
hc-bi & ia-cg & bg-ah\\
bf-ce & cd-af & ae-db
\end{array} \right)= \left(\mathbf{F}^{T}\right)^{-1}
\end{align} 
which is seen to be equivalent to the inverse transpose of the real space deformation gradient, $\mathbf{F}$.

\end{document}